\documentclass[amssymb,amsmath,twocolumn,aps]{revtex4}

\usepackage[dvipdfm]{graphicx}
\usepackage{latexsym,mathrsfs}
\usepackage{dcolumn}
\usepackage{bm}
\usepackage{graphicx}
\usepackage{txfonts}
\usepackage{natbib}
\def \ve#1{\mbox{\boldmath $#1$}}

\begin{document}

\title{
How is the magnetic reconnection derived from magnetohydrodynamics equations? \\
}
\author{Tohru Tashiro}
\email{tashiro@cosmos.phys.ocha.ac.jp}   
\author{Wakako Kakuta}
\affiliation{Department of Physics, Ochanomizu University, 2-1-1 Ohtsuka, Bunkyo, Tokyo 112-8610, Japan}
\date{\today}

\begin{abstract}

We clarify how magnetic reconnection can be derived from magnetohydrodynamics (MHD) equations in a way that is easily understandable to university students. The essential mechanism governing the time evolution of the magnetic field is diffusion dynamics. The magnetic field is represented by two components. It is clarified that the diffusion of a component causes a generation of another component that is initially zero and, accordingly, that the magnetic force lines are reconnected. For this reconnection to occur correctly, the initial magnetic field must be directed oppositely in the two regions, e.g., $y>0$ and $y<0$; must be concave (convex) for $y>0$ ($y<0$); and must be saturated for $y$ far from the $x$ axis, which would indicate the existence of the current sheet. It will be clear that our comprehension based on diffusion runs parallel to the common qualitative explanation about the magnetic reconnection.

\end{abstract}

\maketitle

\section{Introduction}

A solar flare is the most violent explosion in the solar system. It occurs over the Sun's surface, resulting in the brightening of electromagnetic waves over a wide wavelength range and the release of about $10^{22}\mbox{--}10^{25}$ joule of energy within about $10^2\mbox{--}10^3$ s\cite{Yokoyama10}. 

How such a large amount of energy is released in such a short time has puzzled scientists for many years.
One of the epoch-making approaches to this problem is to exchange the magnetic energy into kinetic energy and the heat of plasma by reconnecting magnetic field lines. This is referred to as {\em magnetic reconnection}.
The time evolution of the magnetic field is determined by a magnetic induction equation that includes 
two contrary effects, ``freezing'' and diffusion of the magnetic field, which is explained in detail later.
Diffusion of the magnetic field is caused by Joule dissipation.
If the electric resistivity is low, the magnetic field is ``frozen'' into the plasma.
However, if an anomalous resistivity is induced for some reason and the diffusion effect becomes dominant, the freezing of the magnetic field ceases and its structure can be altered.

In other words, the diffusion causes structural change of the magnetic field.
However, the time for the release of the magnetic energy is quite long. In the case of the solar flare, the diffusion constant $\eta$ is about $1 \ \mbox{m}^2/\mbox{s}$ and the typical scale of length for the energy release, which we shall denote by $L_0$, is $10^8 \ \mbox{m}$. Therefore, the time can be estimated as ${L_0}^2/\eta\sim 10^{16} \ \mbox{s} \sim 10^{8} \ \mbox{yr}$, which is not comparable with the observed release time\cite{Priest}.
Hence, we can explain the heating of plasma and the release of a large amount of energy by considering only the time evolution of the magnetic field, but cannot explain the release time.

For resolving  this problem, Dungey proposed a model considering the electromagnetic field caused by the motion of the plasma and, thus, using magnetohydrodynamics (MHD) with negligible gas pressure. He showed that a current sheet forms between opposite directions of the magnetic field\cite{Dungey53,Priest}.
Dungey was also the first to suggest that the configuration of the magnetic field can change. 
Subsequently, Sweet and Parker expanded on Dungey's model by considering the in/out flow of the plasma and using MHD equations; this model is referred to as {\em Sweet-Parker's model}\cite{Sweet58a,Sweet58b,Parker57,Priest}.
According to their model, a plasma flowing into the current sheet with a velocity $v_{i}$ flows out along the sheet with the Alfv\'{e}n velocity $V_A$. The velocity $v_{i}$ satisfies the following relation.
\begin{equation}
v_i = {R_m}^{-1/2}V_A \ ,
\end{equation}
where ${R_m}$ denotes the magnetic Reynolds number, which can be expressed as
\begin{equation}
R_m = \frac{L_0V_A}{\eta} \ .
\end{equation}
The magnetic energy moves with velocity $v_{i}$, and the time with which the energy is released over a length $L_0$ is given by
\begin{equation}
\frac{L_0}{v_i} = \frac{{L_0}^{3/2}}{\sqrt{\eta V_A}} \ .
\end{equation}
If we use $10^8 \ \mbox{m}$ and $1 \ \mbox{m}^2/\mbox{s}$ as values for $L_0$ and $\eta$, respectively, and estimate the Alfv\'{e}n velocity as $10^{8} \ \mbox{m}/\mbox{s}$, the energy release time becomes $10^8 \ \mbox{s}$.
Therefore, we can find that the time is much shorter than that for the model considering only diffusion dynamics. However, this time is still longer than what is observed.
Petschek reduced the reconnection region in order to resolve this discrepancy, resulting in the release time of $10^2\mbox{--}10^3 \ \mbox{s}$, which is closer to the observed one\cite{Petschek64}. 

\begin{figure}[h]
  \begin{center}
    \includegraphics[scale=.32]{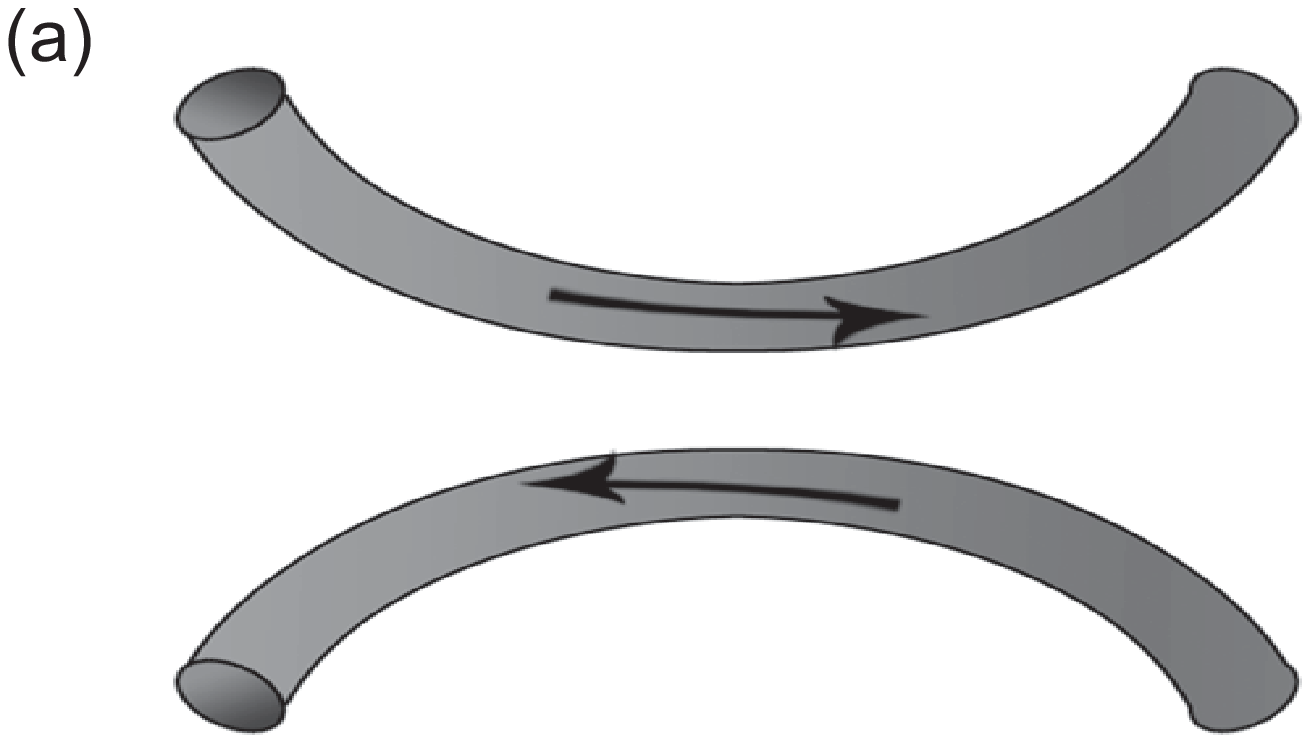}
    \mbox{}
    \includegraphics[scale=.32]{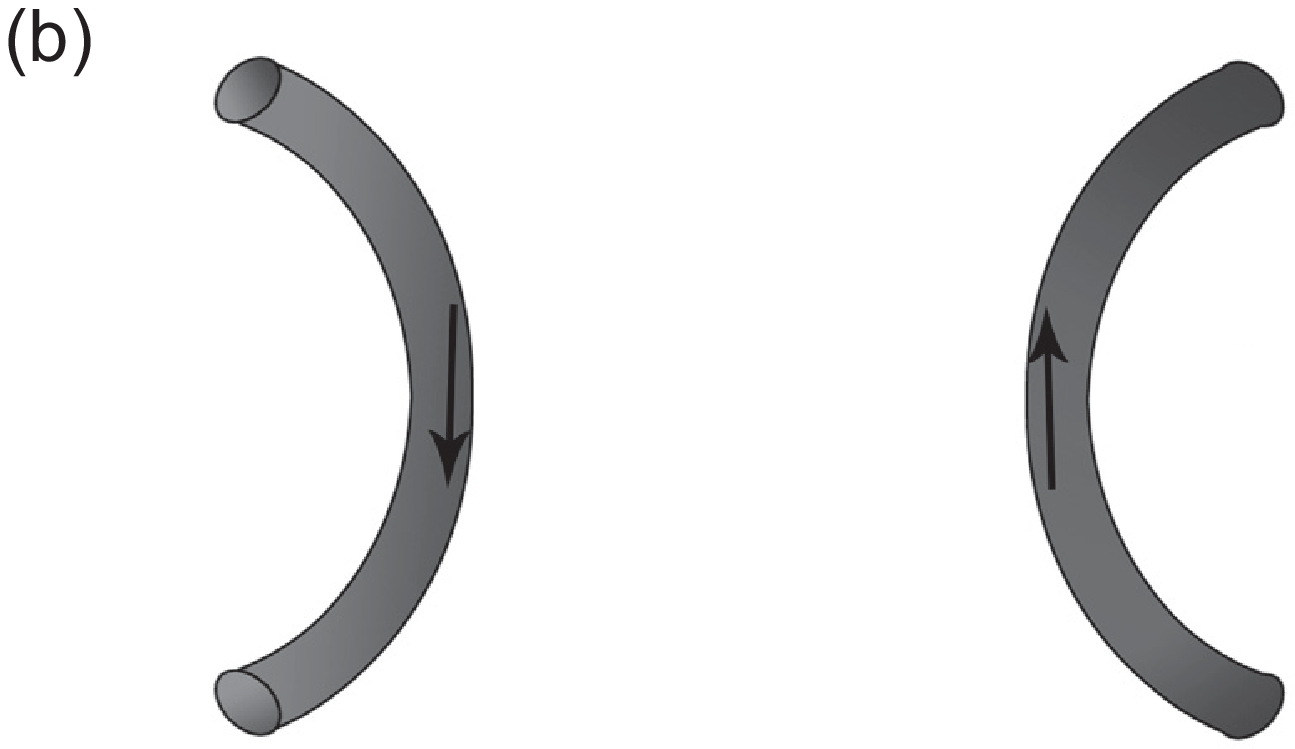}
    \caption{\label{fig:kakuta}Illustration of magnetic reconnection by using an analogy between magnetic field lines and rubber bands. Enormous energy is stored in the extended magnetic field lines shown in (a). The occurrence of dissipation makes the magnetic force lines reconnect, as in (b), resulting in the release of stored energy.}
  \end{center}
\end{figure}

As mentioned above, the theory of magnetic reconnection has been developed. However, the concept used in describing the time evolution of the magnetic field is that of diffusion. Hence, the question arises as to why diffusion produces a structural change of the magnetic field.
Of course, we can often find a qualitative explanation by drawing analogy between magnetic field lines and rubber bands: the energy of short connected magnetic field lines is lower than that of extended magnetic field lines, as is shown in Fig.~\ref{fig:kakuta}. However, this analogy does not account for the aspect of diffusion dynamics.
One of ways to determine that magnetic reconnection is caused by diffusion is to calculate the magnetic induction equation and other MHD equations numerically. However, such a calculation would be required to be done over a complex computer program using an expensive, high-specification machine.
In this paper, we demonstrate how the magnetic field reconnects not through such an advanced technical process, but through a method that can be easily understood by university students who are familiar with diffusion phenomena. 

This paper is organized as follows. First, we derive the equations governing the time evolution of magnetic and velocity fields and define their initial and boundary conditions in Sec.~\ref{FE}. Next, the diffusion phenomena are revisited in Sec.~\ref{diff}, after which we demonstrate how the magnetic field lines are reconnected using the basic knowledge of diffusion in Sec.~\ref{main}. Finally, in Sec.~\ref{CR}, we conclude this paper and indicate the equality of our comprehension and the common qualitative explanation. 

\section{Fundamental equations}
\label{FE}

In this section, we enumerate the equations governing the time evolution of magnetic and velocity fields. For simplicity, these fields are assumed to exist only on the $x\mbox{--}y$ plane.

The time evolution of the magnetic field, $\ve{B}$, is expressed by the following magnetic induction equation.
\begin{equation}
\frac{\partial \ve{B}}{\partial t} = \ve{\nabla}\times(\ve{v}\times\ve{B}) + \eta\triangle\ve{B} \ ,
\label{MIE}
\end{equation}
where $\ve{v}$ is the velocity field of the plasma and $\eta$ is the magnetic diffusivity, which can be defined by using the electric resistivity, $\eta_e$, and the magnetic permeability, $\mu$, as $\eta={\eta_e}/{\mu}$. The right hand side has two terms: the first represents the effect of the freezing of the magnetic field into the plasma, and the second represents the diffusion. For the case with $\eta=0$, i.e., the case of a perfect conductor, the magnetic field moves together with the plasma\cite{Priest,Landau}. 
This magnetic induction equation can be derived by incorporating Faraday's law
\begin{equation}
\frac{\partial \ve{B}}{\partial t} = -\ve{\nabla}\times\ve{E} \ ,
\end{equation}
the electric field of the Ohm's law
\begin{equation}
\eta_e\ve{j} = \ve{E} + \ve{v}\times\ve{B} \ ,
\label{eq:Ohm}
\end{equation}
and Amp\`ere's law
\begin{equation}
\ve{\nabla}\times\ve{B} = \mu\ve{j} \ ,
\end{equation}
in which $\ve{j}$ is the current density. See Appendix for the details.

Next, the equation of motion of the plasma that is considered as a perfect incompressible fluid is
\begin{equation}
\rho_m\left\{\frac{\partial\ve{v}}{\partial t}+(\ve{v}\cdot\ve{\nabla})\ve{v}\right\} = -\ve{\nabla}p + \ve{j}\times\ve{B} \ ,
\label{euler}
\end{equation}
where $\rho_m$ and $p$ represent the mass density and the pressure of plasma, respectively.
The above two equations, Eqs~(\ref{MIE}) and (\ref{euler}), are sufficient for us to understand the magnetic reconnection.

We show the change in the electric resistivity, $\eta_e$, with time, which is how the structural change of the magnetic field starts, as follows: before $t=0$, the value is zero everywhere; subsequently, it increases drastically only over the region $-L< x< L$ and $-w< y< w$, which we call the {\em reconnection region}. Hence, the magnetic field ``frozen'' into the plasma starts to ``melt'' at $t=0$. Then, we set the change in magnetic diffusivity, $\eta$, in this region as follows:
\begin{equation}
\eta = \left\{
\begin{array}{cc}
0                 & (t<   0) \\
\eta^*  \ (\gg 1) & (t\ge 0)
\end{array}
\right. \ .
\end{equation}
In addition, we shall call the other region, $|x|\ge L$ or $|y|\ge w$, the {\em frozen region}.

The initial magnetic field, $\ve{B}(\ve{r},t\le0)$, is chosen to be parallel with the $x$ axis at $y>0$ and antiparallel with it at $y<0$. Thus, we adopt an odd function as the magnetic field. Here, we use the following simple expression
\begin{align}
\ve{B}(\ve{r},t\le0) &= (B_x(\ve{r},t\le0),B_y(\ve{r},t\le0),0) \nonumber \\  
                     &= B_0(\tanh ay,0,0) \ ,
\label{eq:B0}
\end{align}
where $a$ is some constant whose dimension is the inverse of length. 

From Amp\`ere's law, the current density can be obtained as
\begin{align}
\ve{j}(\ve{r},t\le0) &= \frac{1}{\mu}\ve{\nabla}\times\ve{B}(\ve{r},t\le0) \nonumber \\
                     &= \frac{1}{\mu}(0,0,\partial_xB_y(\ve{r},t\le0)-\partial_yB_x(\ve{r},t\le0)) \nonumber \\
                     &= -\frac{aB_0}{\mu}(0,0,{\rm sech}^2ay) \ . \label{currentdens}
\end{align}
This current density localizes around the $x$ axis, that is, the current sheet. The existence of this current sheet between oppositely directed magnetic fields is important in the magnetic reconnection.
Therefore, we choose the hyperbolic tangent among odd functions: the derivative of the initial $B_x$ with respect to $y$ must localize around the $x$ axis. In other words, this initial magnetic field must be concave (convex) at $y>0$ ($y<0$) and be saturated for $y$ far from the $x$ axis.

In this paper, we regard the plasma as a perfect conductor with no resistivity until $t=0$, i.e., $\eta=\eta_e=0$. The velocity field must be zero in order to keep this magnetic field static:
\begin{align}
\ve{v}(\ve{r},t\le0) &= (v_x(\ve{r},t\le0),v_y(\ve{r},t\le0),v_z(\ve{r},t\le0)) \nonumber \\
                     &= (0,0,0)  \ . \label{eq:v0}
\end{align}
This is because the plasma and the magnetic field move together for the case with $\eta=0$.

For the velocity field to maintain this state, the following relation, obtained from Eq.~(\ref{euler}), must be satisfied.
\begin{equation}
-\ve{\nabla}p = - \ve{j}\times\ve{B}
\label{kikko}
\end{equation}
From Eq.~(\ref{currentdens}), the right hand side of Eq.~(\ref{kikko}) becomes
\begin{align}
\lefteqn{-\ve{j}(\ve{r},t\le0)\times\ve{B}(\ve{r},t\le0)} \nonumber \\
 &= ({j}_z(\ve{r},t\le0)B_y(\ve{r},t\le0) , -{j}_z(\ve{r},t\le0)B_x(\ve{r},t\le0) , 0) \nonumber \\ 
 &= \frac{a{B_0}^2}{\mu}(0,{\rm sech}^2 ay \, \tanh ay,0) \ .
\end{align}
Hence, we obtain the pressure gradient, 
\begin{equation}
-\ve{\nabla}p = \frac{a{B_0}^2}{\mu}(0,{\rm sech}^2 ay \, \tanh ay,0) \ .
\label{eq:pgrad}
\end{equation}
These magnetic and velocity fields do not change in the frozen region after $t=0$.

\section{Time evolution of magnetic and velocity fields}

\subsection{diffusion}
\label{diff}

In this subsection, we revisit the diffusion dynamics. Although diffusion is not the primary topic of this paper, the main equation governing the time evolution of the magnetic field is the diffusion equation, as shown in Eq.~(\ref{MIE}).

\begin{figure}[h]
  \begin{center}
    \includegraphics[scale=.52]{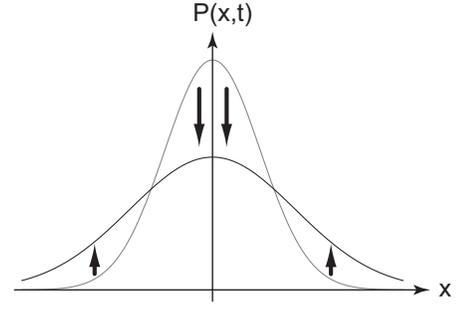}
    \caption{\label{fig:diffusion}Time evolution of the diffusion equation, Eq.~(\ref{eq:diffusion}), from an initial state (gray line) to a later state (black line). The concave region of $P(x,t)$ decreases, and the convex one swells up.}
  \end{center}
\end{figure}

The diffusion equation in one dimension is 
\begin{equation}
\partial_t P(x,t) = D\partial_x^2 P(x,t) \ ,
\label{eq:diffusion}
\end{equation}
in which $D$ is the diffusion constant. 
We can interpret this equation as follows: $P(x,t)$ increases (decreases) with time if the second derivative with respect to the spatial variable is positive (negative).
$P(x,t)$ is convex (concave) at a region where the second derivative is positive (negative).
Therefore, a picture of the diffusion dynamics is a rise of the convex region and a fall of the concave one, as is shown in Fig.~\ref{fig:diffusion}.

\subsection{magnetic reconnection}
\label{main}

Here, we investigate how the magnetic field changes for a small $t$ by the occurrence of dissipation at $t=0$.

From the values of the magnetic and velocity fields at $t\le0$, shown as in the previous section, one can obtain
\begin{equation}
\left.{\partial_t \ve{B}}\right|_{t=0} = {\eta^*}\triangle\ve{B}
\end{equation}
and
\begin{equation}
\left.{\partial_t\ve{v}}\right|_{t=0} = \ve{0} \ .
\end{equation}
From these equations, it is clarified that the magnetic field evolves with time earlier than the velocity field.
Hence, let us consider the time evolution of the magnetic field maintaining the value of the velocity field as that in Eq.~(\ref{eq:v0}), and then, let us investigate how this time evolution affects the zero velocity field. 

Because we can regard $B_y$ at a small $t$ as being nearly equal to zero, magnetic induction equations at the reconnection region become
\begin{equation}
 {\partial_t B_x(\ve{r},t)} = -{\partial_y}v_y(\ve{r},t)B_x(\ve{r},t) + {\eta^*}\triangle B_x(\ve{r},t) 
\label{IE:Bx}
\end{equation}
and
\begin{equation}
 {\partial_t B_y(\ve{r},t)} = {\partial_x}v_y(\ve{r},t)B_x(\ve{r},t)  \ .\label{IE:By}
\end{equation}
Moreover, from Eqs.~(\ref{eq:B0}) and (\ref{eq:v0}), we can regard $B_x$ and $v_y$ at a small $t$ as
\begin{equation}
B_x(\ve{r},t)\simeq B_0\tanh ay
\label{eq:Bx0}
\end{equation}
and
\begin{equation}
v_y(\ve{r},t)\simeq 0 \ . 
\label{eq:vy0}
\end{equation}
Therefore, the right hand side of Eq.~(\ref{IE:By}) can be considered to be zero at such a time stage. That is to say, $B_x$ changes with time earliest among the components of the magnetic field. 

By solving Eq.~(\ref{IE:Bx}), we obtain the behavior of $B_x$ only at $y>0$, as the initial $B_x$ is odd under the transformation $y\rightarrow-y$ and this equation is invariant under transformations $B_x\rightarrow-B_x$ and $y\rightarrow-y$. \footnote{For $y<0$, $B_x(x,y,t)$ is always equal to $-B_x(x,|y|,t)$} 
Therefore, the relation $B_x(x,0,t) = 0$ is always satisfied.

\begin{figure}[h]
  \begin{center}
    \includegraphics[scale=.62]{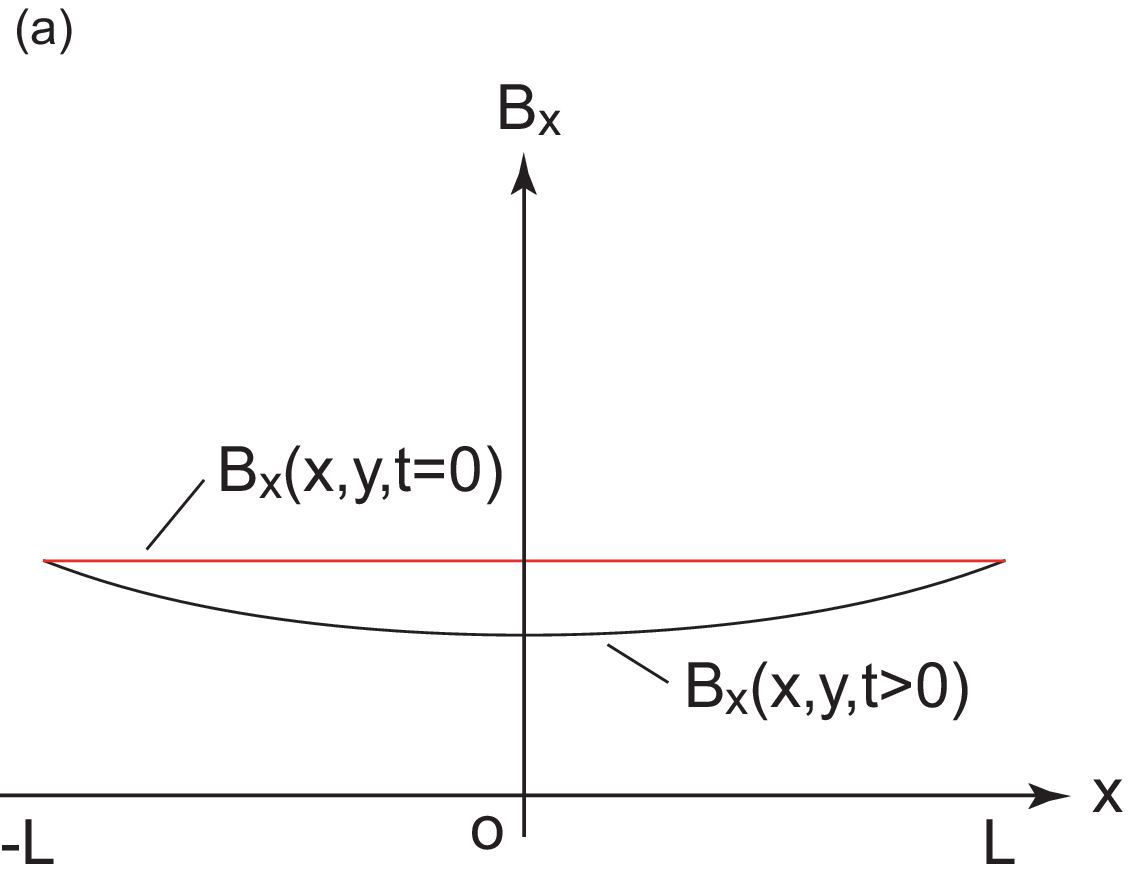}
    \mbox{}
    \includegraphics[scale=.62]{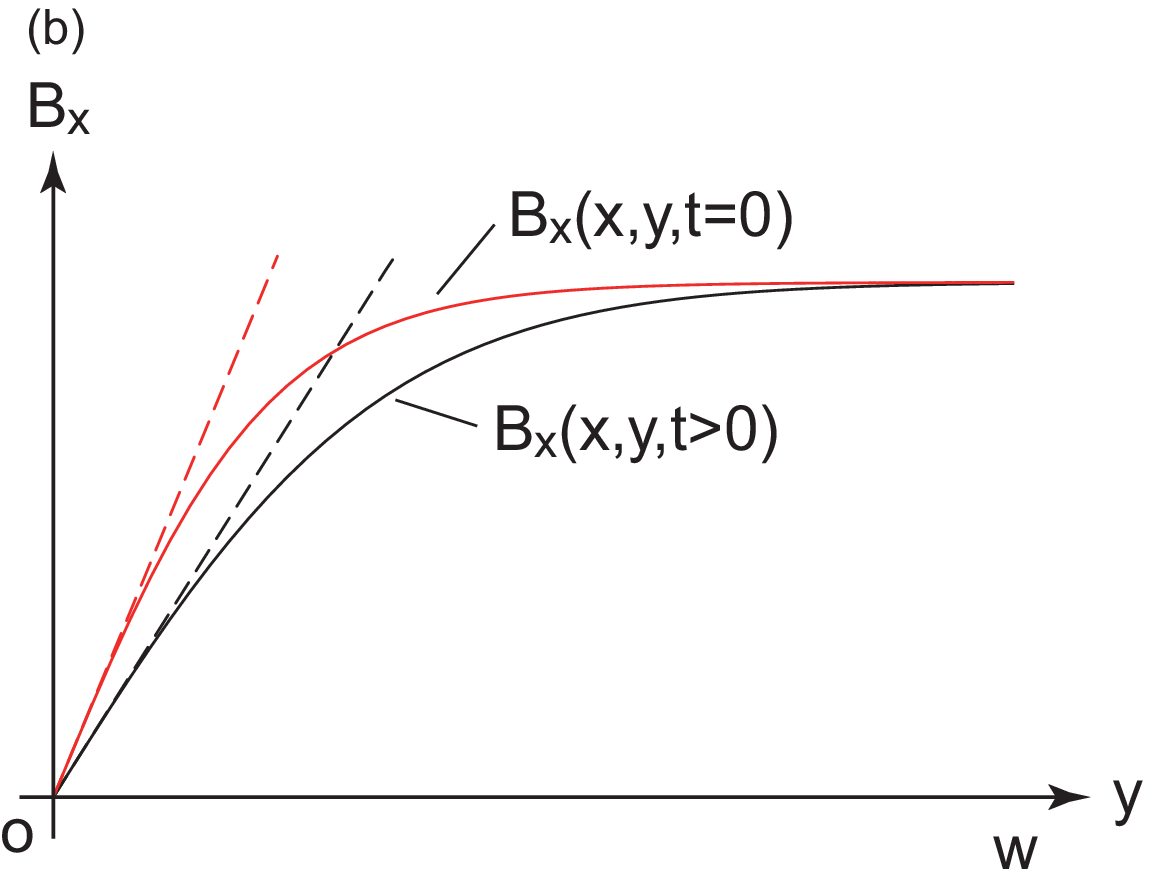}
    \caption{\label{fig:Bx}(color online) Time evolution of $B_x$ and its derivative with respect to $y$ in the reconnection region. The red and black curves respectively represent $B_x(x,y,t>0)$ and $B_x(x,y,t=0)$, that is, $B_0\tanh ay$. The dashed lines in (b) are the derivatives of the curves at $y=0$. From both figures, it is clarified that a denting effect of the diffusion process works. In particular, we can find from (b) that the slope of the red dashed curve denoting $\partial_yB_x(x,y,t=0)$ is larger than that of the black curve denoting $\partial_yB_x(x,y,t>0)$.}
  \end{center}
\end{figure}

Equations (\ref{IE:Bx}) and (\ref{eq:vy0}) together compromise the exact diffusion equation. 
Because the initial magnetic field, $B_x(\ve{r},0)=B_0\tanh ay$, is concave for $y>0$, the effect that results in denting of the convex region works in the same way as that explained in Sec.~\ref{diff}.
As mentioned in the above paragraph, however, $B_x$ is always zero at $y=0$. Moreover, the value of the magnetic field is fixed at $y=w$ and $x=\pm L$, as these are boundaries between the reconnection and frozen regions. Therefore, the denting is remarkable at the center of the reconnection region with $y>0$, as shown in Fig.~\ref{fig:Bx}; this denting is indicated by
\begin{equation}
  {\partial_x B_x} \left\{
  \begin{array}{ll}
    >0 & (\mbox{for} \ x>0) \\
    <0 & (\mbox{for} \ x<0)
  \end{array}
  \right. \ . \label{ahoMM}
\end{equation}
This absolute value increases with the distance from the $y$ axis, which can be seen easily from Fig.~\ref{fig:Bx}(a).

Because of the time evolution of $B_x$,  
the $z$ component of the current density, ${j}_z$, which is described by using the derivatives of the magnetic field as
\begin{equation}
{j}_z(\ve{r},t)= \frac{1}{\mu}\left\{\partial_xB_y(\ve{r},t)-\partial_yB_x(\ve{r},t)\right\} \ ,
\end{equation}
changes with time in the following way: $\partial_yB_x$  becomes smaller than its initial value, i.e., $aB_0\,{\rm sech}^2\,ay$, particularly at a small $y$ because of the denting effect of diffusion (See Fig.~\ref{fig:Bx}(b)).
Therefore, the absolute value of $j_z$ localizing in the vicinity of the $x$ axis at $t=0$ diminishes. The following qualitative explanation may help us understand this result easily: The occurrence of dissipation reduces the current density.
Hence, the balance between the pressure gradient and the force from the magnetic field described by Eq.~(\ref{kikko}) breaks down; the time evolution of the velocity field is thus governed only by the pressure gradient. Because this gradient is shown as Eq.~(\ref{eq:pgrad}), only the $y$ component of the velocity field has the following value, which does not depend on $x$:
\begin{equation}
  v_y \left\{
  \begin{array}{ll}
    >0 & (\mbox{for} \ y>0) \\
    <0 & (\mbox{for} \ y<0)
  \end{array}
  \right.  \ .\label{eq:vygrow}
\end{equation}
This absolute value becomes smaller with the distance from the $x$ axis.
The reason is as follows. 
The absolute value of the current density is originally small at a region far from the $x$ axis, and hence, the pressure gradient balanced with the force caused by the current density and the magnetic field is also small at this region.

As we have seen, ${\partial_x B_x}$ and $v_y$ have non-zero values after $t=0$ owing to the diffusion dynamics in spite of the fact that these initial values are zero.
From Eq.~(\ref{IE:By}), the change of $B_y$ with time is determined by the product of ${\partial_x B_x}$ and $v_y$, since $v_y$ is independent of $x$. In short, $B_y$ starts to evolve with time since the right hand side of Eq.~(\ref{IE:By}) is no longer zero. Considering $B_y(\ve{r},0)=0$, Eq.~(\ref{ahoMM}), and $v_y>0$ for $y>0$, one can reach the following conclusion:
\begin{equation}
  B_y \left\{
  \begin{array}{ll}
    >0 & (\mbox{for} \ x>0) \\
    <0 & (\mbox{for} \ x<0)
  \end{array}
  \right. \ . \label{eq:Bygrow}
\end{equation}
This absolute value increases with distance from the $y$ axis.
On the other hand, this value decreases with distance from the $x$ axis because of the behavior of $v_y$ as mentioned before.

\begin{figure}[h]
  \begin{center}
    \includegraphics[scale=.37]{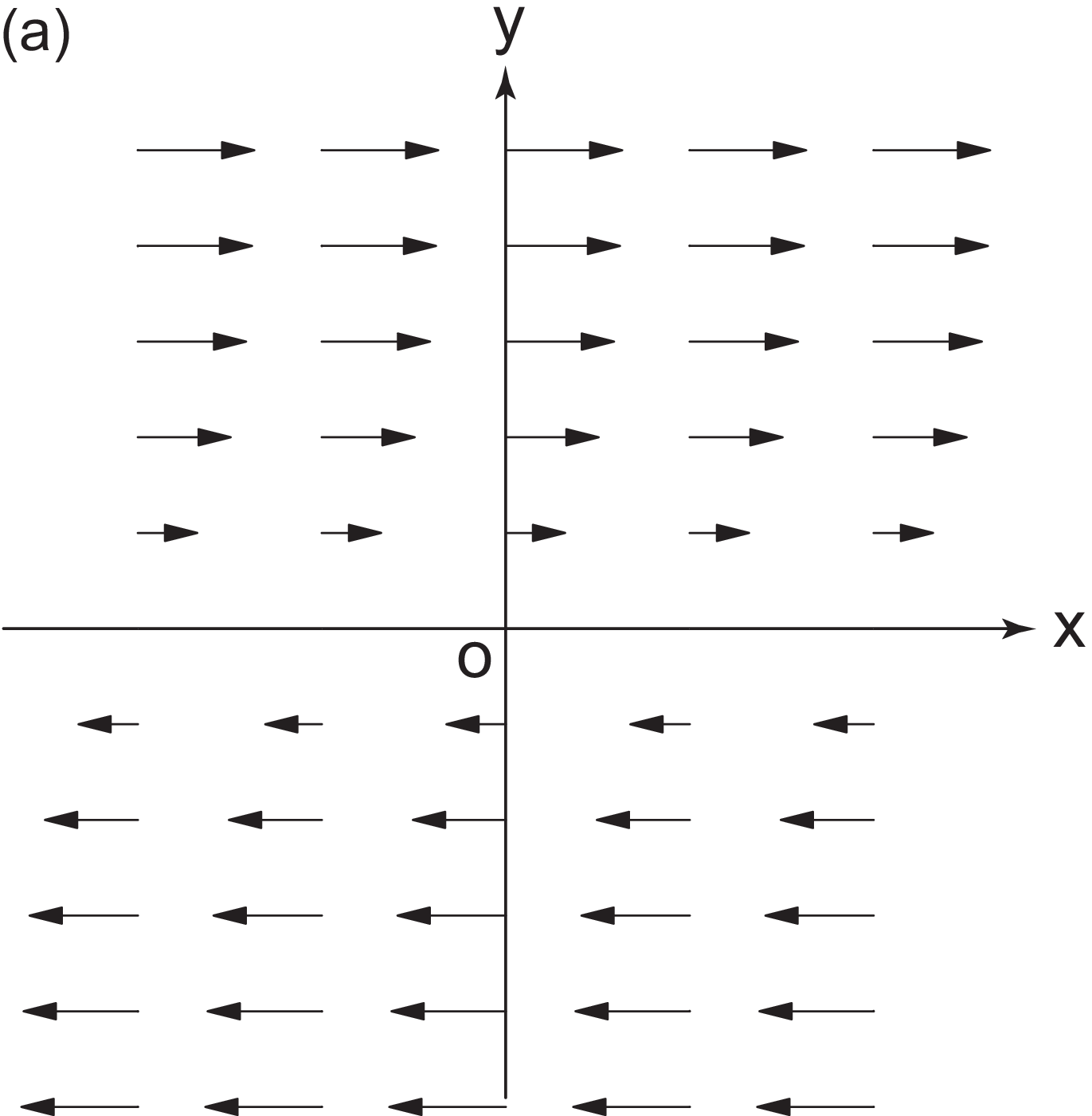}
\mbox{}
    \includegraphics[scale=.37]{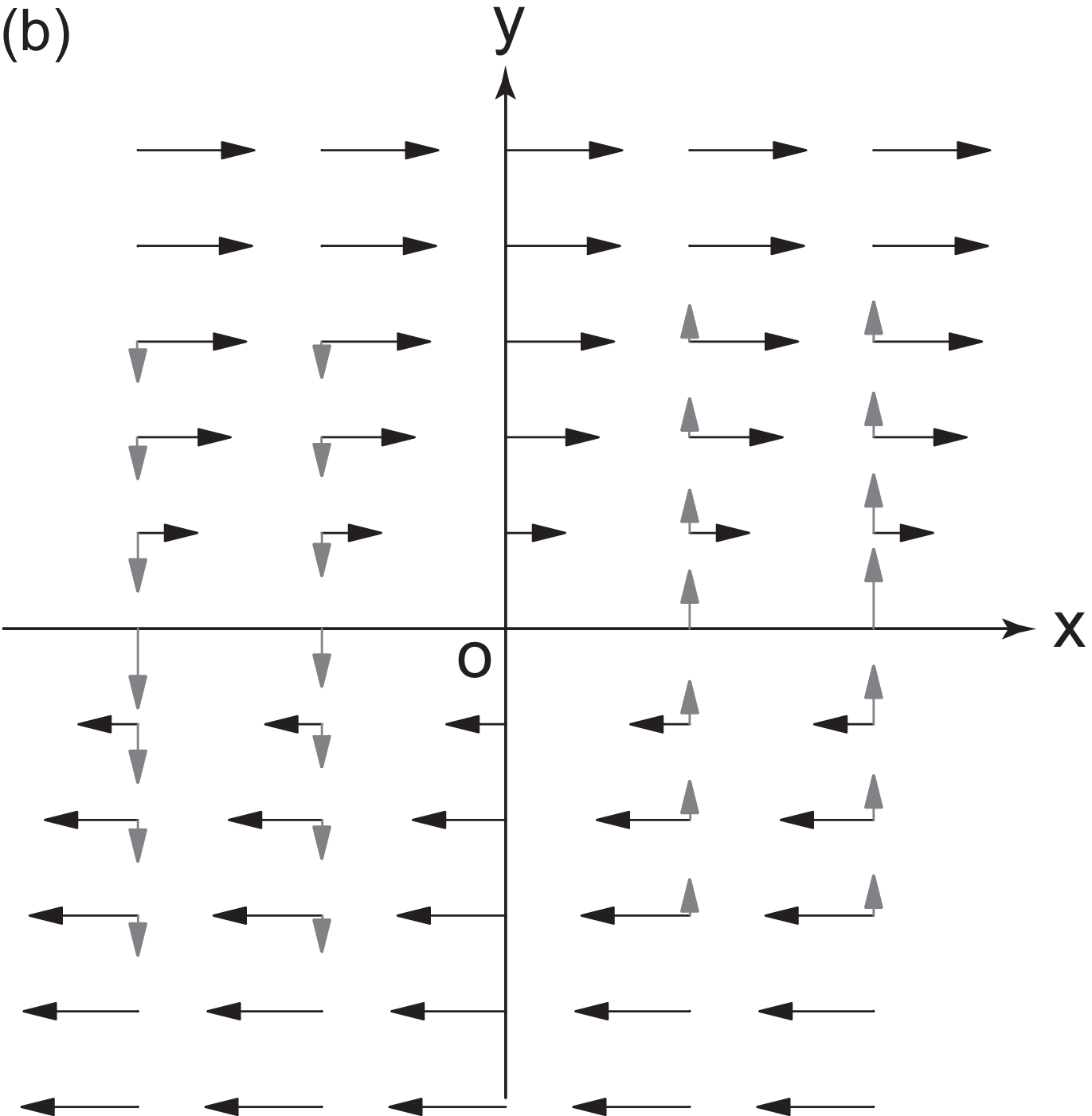}
\mbox{}
    \includegraphics[scale=.37]{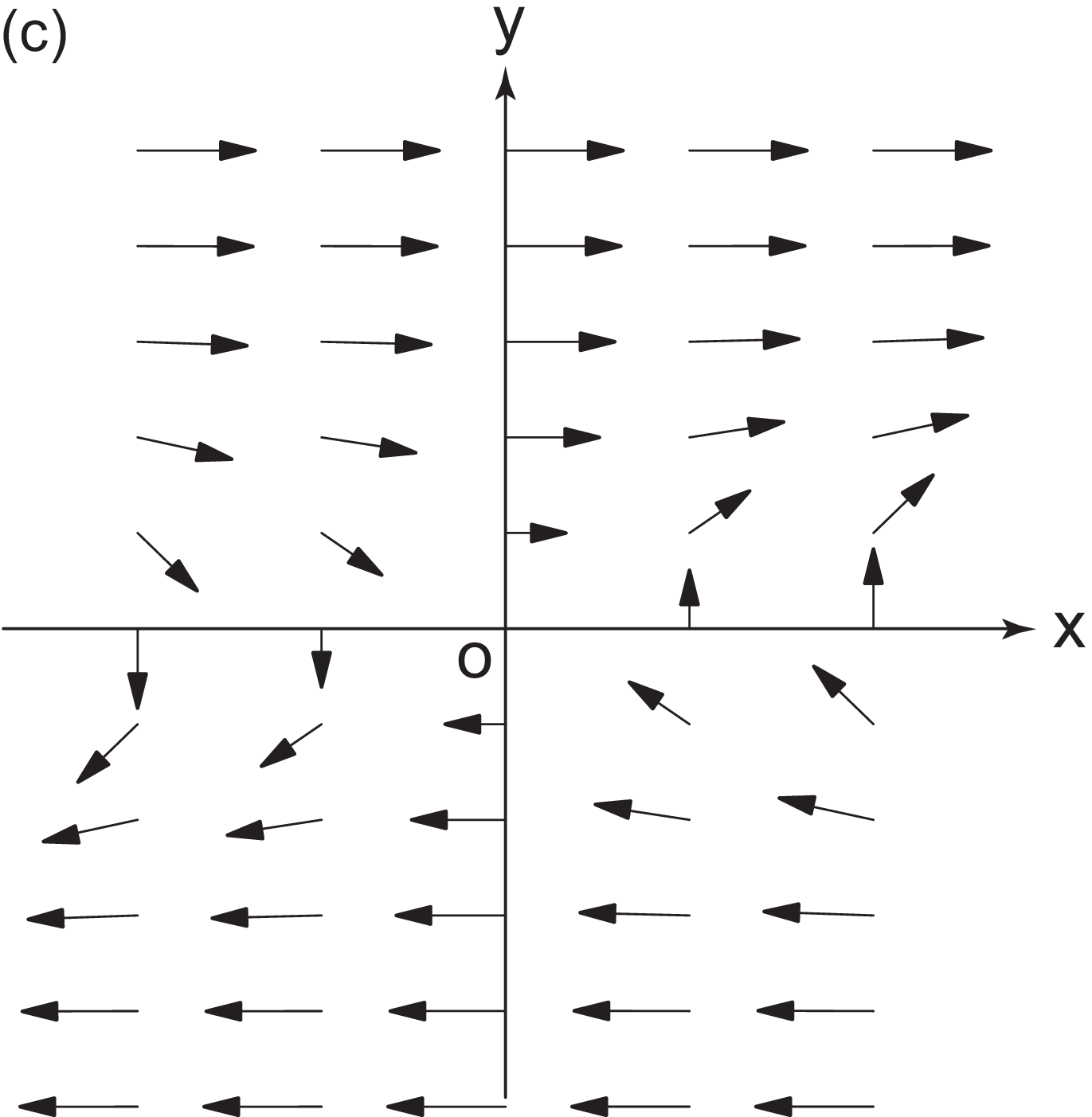}
    \caption{\label{fig:magnetic}Illustration of magnetic reconnection based on our comprehension: (a) initial magnetic field described by Eq.~(\ref{eq:B0}) where $B_y=0$, (b) diffusion of $B_x$ causes emergence of $B_y$, denoted by gray arrows, and (c) the composed magnetic field.}
  \end{center}
\end{figure}

Hence, we can understand that the magnetic field directed parallelly or antiparallelly along the $x$ axis at $t=0$ starts to reconnect owing to the occurrence of $B_y$, which is described in Fig.~\ref{fig:magnetic}.  

\section{Concluding remarks}
\label{CR}

We have clarified how the diffusion gives rise to the magnetic reconnection. The magnetic induction equation has two components. First, the $x$ component, $B_x$, diffuses, and then, the $y$ component, $B_y$, which is initially zero, is generated by the time evolution of $B_x$.
For the magnetic field lines to be reconnected correctly, the initial $B_x$ must be concave (convex) at $y>0$ ($y<0$) and must be saturated for $y$ far from the $x$ axis to such an extent as $1/a$ as in this paper; this behavior would indicate that the current density is localized around the $x$ axis, i.e., the current sheet.
This current sheet is also necessary in the qualitative explanation. If an anomalous resistivity occurs, the current dissipates and the magnetic field starts to melt. As a result, enormous energy breaks out through the Joule dissipation and magnetic reconnection is generated.
The common qualitative explanation of the magnetic reconnection using the current sheet is connected to our comprehension based on diffusion in this paper.
Finally, we emphasize that if we adopt other concave (convex) functions of $y$ at $y>0$ ($y<0$), which are saturated as the initial magnetic field, magnetic reconnection can occur.

\appendix

\section{Magnetic induction equation}

By substituting the electric field of Ohm's law,
\begin{equation}
\eta_e\ve{j} = \ve{E} + \ve{v}\times\ve{B} \ ,
\end{equation}
into the left-hand side of Faraday's law,
\begin{equation}
\frac{\partial \ve{B}}{\partial t} = -\ve{\nabla}\times\ve{E} \ ,
\end{equation}
we can obtain
\begin{equation}
\frac{\partial \ve{B}}{\partial t} = \ve{\nabla}\times(\ve{v}\times\ve{B}) -{\eta_e}\ve{\nabla}\times\ve{j} \ .
\label{eq:app1}
\end{equation}

Amp\`ere's law, i.e., 
\begin{equation}
\ve{\nabla}\times\ve{B} = \mu\ve{j}
\end{equation}
relates the current density to the magnetic field. Therefore, equation (\ref{eq:app1}) becomes
\begin{equation}
\frac{\partial \ve{B}}{\partial t} = \ve{\nabla}\times(\ve{v}\times\ve{B}) -\frac{\eta_e}{\mu}\ve{\nabla}\times(\ve{\nabla}\times\ve{B}) \ .
\end{equation}

For vector $\ve{a}$, the following relation is satisfied.
\begin{equation}
\ve{\nabla}\times(\ve{\nabla}\times\ve{a}) = \ve{\nabla}(\ve{\nabla}\cdot\ve{a}) - \triangle\ve{a} 
\end{equation}
With regard to the magnetic field, $\ve{\nabla}\cdot\ve{B}=0$ is satisfied, as there is no magnetic monopole. Thus, 
\begin{equation}
\ve{\nabla}\times(\ve{\nabla}\times\ve{B}) = - \triangle\ve{B} \ ,
\end{equation}
and so, the induction equation is obtained as follows:
\begin{equation}
\frac{\partial \ve{B}}{\partial t} = \ve{\nabla}\times(\ve{v}\times\ve{B}) + \frac{\eta_e}{\mu}\triangle\ve{B} \ .
\end{equation}

\begin{acknowledgments}
We would like to thank Hiroe Minagawa, Prof. Morikawa Masahiro, and members of Astrophysics Laboratory at Ochanomizu University for the extensive discussions.
\end{acknowledgments}

\end{document}